\documentstyle[prb,aps,multicol,epsfig]{revtex}

\begin{document}

\title{Proposal for a Quantum Hall  Pump}

\author{Steven H. Simon}
\address{Lucent Technologies Bell Labs, Murray Hill NJ 07974} 

\maketitle

\begin{abstract} 
  A device is proposed that is similar in spirit to the electron
  turnstile except that it operates within a quantum Hall fluid.  In
  the integer quantum Hall regime, this device pumps an integer number
  of electrons per cycle.  In the fractional regime, it pumps an
  integer number of fractionally charged quasiparticles per cycle.  It
  is proposed that such a device can make an accurate measurement of
  the charge of the quantum Hall effect quasiparticles.
\end{abstract}

\begin{multicols}{2}

\narrowtext

The basic idea of a parametric pump is that some parameters of a
system are varied slowly and periodically such that after each full
cycle the system returns to its initial state with the net effect
being that some amount of a fluid is transferred from a source to a
drain.  There are many examples of such pumps in a very wide range of
contexts --- from the human heart to a firemans' bucket brigade.  Over
the past few years there has been increasing interest in parametric
pumping of charge in mesoscopic systems both
theoretically\cite{Theory,Brouwer} and
experimentally\cite{Experiment1,Experiment2,Standard}.  One
particularly interesting example of a parametric pump is the electron
turnstile -- a device that transfers a single electron per cycle from
a source to a drain.  Such devices seem quite promising as
metrological current and capacitance
standards\cite{Experiment2,Standard}.  In this paper I propose a
device very similar to the electron turnstile that operates in the
quantum Hall regime.  Similar to the electron turnstile, when operated
adiabatically at low temperature in the integer quantum Hall regime,
the number of electrons pumped in a single cycle is quantized.
However, in the fractional quantum Hall regime, it is an integer
number of {\it fractionally charged quasiparticles} that is pumped in
each cycle.  Thus, this device has the potential to make measurements
of the fractional charge of quantum Hall quasiparticles.

\begin{figure}[htbp]
  \hspace*{20pt} \epsfig{file=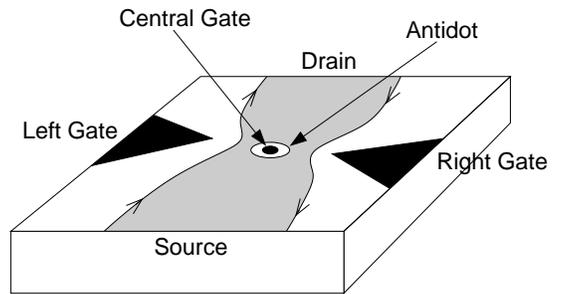,height=1.5in}
  \caption{Cartoon schematic of the proposed quantum Hall pump. 
    The lightly shaded region in the center is quantum Hall fluid.
    The black areas are gates.  Arrows at the edges of the fluid
    indicate edge state propagation direction.  The side gates can
    push the edges of the fluid closer to or further from the central
    antidot.  The small gate in the center can change the size of the
    antidot.}
  \label{fig:exp}
\end{figure}

{\bf Description of the Device:} The structure of the proposed device,
shown schematically in Fig.~\ref{fig:exp}, is quite similar to the
devices used in Refs.~\onlinecite{Sac,Pepper1,Goldman,Pepper2}.

\end{multicols}

\widetext

\begin{figure}[htbp]
\hspace*{15pt}  \epsfig{file=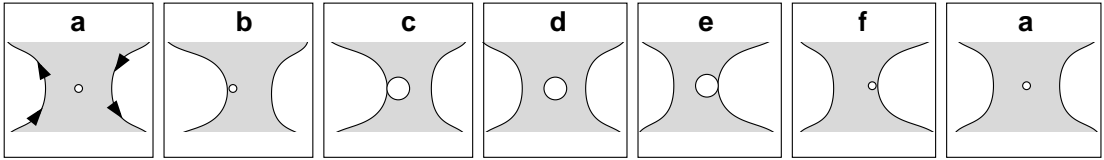,height=1.0in} 
    \caption{A full cycle of pumping (a-b-c-d-e-f-a).   Each frame is
      a top view of the device at a different point in the pumping
      cycle.  In frame (a), the direction of edge state propagation is
      also shown.  This pumping cycle transfers charge from the the
      source (bottom) to the drain (top) at zero applied source-drain
      voltage.  Note that the anti-dot does not connect to both edges
      simultaneously, so at any moment during the cycle the quantized
      Hall fluid (shaded) connects the source to the drain and the
      source-drain conductance is quantized.  Analogous to the
      electron turnstile, the antidot picks up charge (holes) from the 
      left edge, moves over to the right edge, and then releases the
      charge (and then repeats the process).  Since the amount of
      charge carried by the antidot is quantized, so is the resulting
      pumped current per cycle.  In the integer regime, the charge on
      the antidot (and hence the pumped current per cycle) is
      quantized in units of the electron charge, whereas in the
      fractional regime it is quantized in units of the fractionally
      charged quasiparticle.}
    \label{fig:sequence}
\end{figure}

\begin{multicols}{2}

\narrowtext

A full pumping cycle is shown in Fig.~\ref{fig:sequence}.  Throughout
the cycle, the source-drain voltage may be held at zero.  The cycle
can be described as the following steps:

{\bf (a)} Begin in a state where the edges are far from the antidot.
In this state tunneling from the antidot to either the right or left
edge is forbidden.  (I.e., the tunneling amplitude is very close to
zero). \\
{\bf (b)} Move the left edge state close to the antidot (by charging
the left gate negatively) such that the tunneling amplitude between
the left gate and the antidot becomes large (compared to the pumping
frequency). \\
{\bf (c)} Negatively charge the central gate such that the size of the
antidot grows.  Here, as the potential of the central gate increases,
particles (or quasiparticles) that were occupying states near the
edges of the antidot are shifted above the Fermi energy. As they cross
through the Fermi energy, they tunnel out to the left edge (they
cannot tunnel to the right edge because the right edge is insulated
from the dot by a large region of quantum Hall fluid).  \\
  {\bf (d)} Move the left edge state back to its original position far
from the antidot (by uncharging the left gate) such that tunneling
from the antidot to either the right or left edge is once again
forbidden. \\
  {\bf (e)} Move the right edge state close to the antidot (by charging
the right gate negatively) such that the tunneling amplitude between
the right edge and the antidot becomes large. \\
 {\bf (f)} Uncharge the central gate such that antidot becomes smaller.
As the potential on the central gate decreases, the quasiparticles
from the right edge tunnel back to the region near the edges of the
antidot, filling states that were above the Fermi energy. \\ 
{\bf (a)} Move the right edge back far away from the antidot (by
uncharging the right gate) to return the system to the original state.
  
\vspace*{5pt}

Similar to the electron turnstile the charge pumped in this cycle is
given by the difference between the charge on the antidots in
steps (a) and (d).  It is important to note that in stages (a) and
(d), when the tunneling to both edges is turned off, the charge on the
antidot is quantized either in units of the electron charge (in the
integer regime) or in units of the quasiparticle charge (in the
fractional regime).  Thus, we expect that the charge pumped in a cycle
will similarly be quantized, at least at low temperature.  More
rigorous arguments for this quantization will be made below.

\begin{figure}[htbp]
\hspace*{10pt}    \epsfig{file=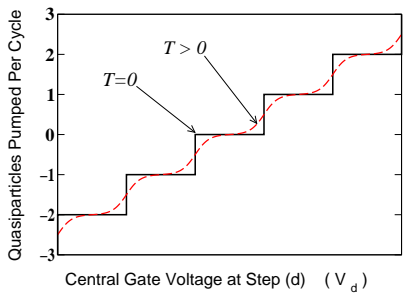,height=1.8in}
    \caption{Pumped charge as a function of central gate voltages $V_d$ 
     at step (d) in the pumping cycle (schematic).  It is assumed that
     the other parameters of the pumping cycle --- and in particular
     the central gate voltage at step (a) are held constant as $V_d$
     is changed.  The solid line is zero temperature, whereas the
     dashed line is finite temperature with $T$ roughly 10\% of the
     single particle addition energy.}
    \label{fig:steps}
\end{figure}

If we then imagine that we fix the central gate voltage at stage (a)
and measure the charge pumped per cycle as a function of the central
gate voltage at stage (d), at zero temperature, we would obtain a
step-like curve, illustrated as the solid line in Fig.~\ref{fig:steps}.

\begin{figure}[htbp]
    \epsfig{file=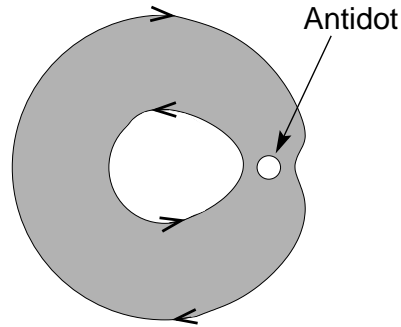,height=2.0in}
    \caption{Quantum Hall Pump  in a Corbino
      geometry} 
    \label{fig:ann}
\end{figure}

{\bf Quantization of Pumping --- Integer Case:} A general approach to
understanding quantized charge pumping is reminiscent of Laughlin's
argument for quantized Hall conductance\cite{Laughlin}.  Consider the
Corbino geometry shown in Fig.~\ref{fig:ann}.  In the integer quantum
Hall regime, at low temperature, the ground state of the system is
unique and gapped at all times in the pumping cycle.  If the
deformation is made adiabatically, the system simply tracks the ground
state\cite{endnote1}.  (``Adiabatic'' here is defined to mean that the
system tracks the ground state).  Thus, at the beginning and end of
the cycle, the system is in the same state and the only net effect is
that an integer number of electrons could have been transferred from
the inside to the outside edge of the annulus (or vice-versa).

For the simple case of non-interacting electrons, one can write the
dynamics in terms of a simple time dependent Schroedinger equation.
This can be integrated explicitly (exactly, or perturbatively) to
demonstrate the quantization of pumped charge as claimed above.  This
explicit approach is useful in that it allows us to study the effects
of nonadiabaticity in detail.  Such a study is a subject of current
research and will be reported elsewhere.

{\bf Fractional Case:} In the case of the fractional quantum Hall
effect, the Laughlin argument must be modified to account for
fractionalization of charge\cite{Stone}.  It now becomes possible to
transfer a single fractionally charged quasiparticle across the
system.  (As usual, increasing the charge on the antidot by a
fractional amount results in the decrease of the charge on the edges
of the system by the same amount being that the bulk is incompressible
and the total charge of the system is conserved).  The argument given
in the above section --- which would seem to require transfer of an
integer number of electrons per cycle --- fails in the fractional Hall
effect case because the ground state becomes $q$-fold
degenerate\cite{Stone} with $q$ a small integer related to the
quasiparticle charge and the filling fraction. For example, for the
simple case of $\nu=p/(2p+1)$, there are $q=2p+1$ degenerate ground
states (and the quasiparticle charge is $e/(2p+1)$).  Because of this
ground state degeneracy, the system need not return to the same ground
state after each pumping period, but may instead cycle through the $q$
ground states.  As a result, it is the number of electrons transferred
across the system in $q$ cycles that is quantized, rather than the
number transferred in a single cycle. Thus, the average charge
transferred in a single cycle is quantized in units of $e/q$, which is
the quasiparticle charge.  Indeed, it is known that adiabatic transfer
of a quasiparticle across such a Corbino system does indeed cycle the
degenerate ground states\cite{Stone}.

Other than this minor modification of the above Laughlin-like
argument, we expect that the same considerations as in the above
integer case will apply for all fractional quantized Hall states.  We
also expect that, as above, the temperature scale at which the
quantization is smeared out is roughly given by the single
quasiparticle addition energy.  For a more detailed calculation, we
expect that chiral Luttinger liquid theory\cite{Lutt} can be used to
calculate the pumped current explicitly.  This, too, is a subject of
current research, and will be reported elsewhere.

{\bf Scattering Matrix Approach:} A rather elegant, more formal,
argument for quantization is based on the scattering matrix approach
to adiabatic parametric pumping\cite{Brouwer}.  In this approach, one
writes the charge pumped in one cycle ($t$ varies from $0$ to $\tau$)
as
\begin{equation}
Q =e \int_0^\tau \frac{dt}{2 \pi} \,
 \sum_{\beta} \sum_{\alpha \in 
  {\mbox{\small source}}} {\rm{Im}}  \left[  S^*_{\alpha \beta}(t) 
\frac{d}{dt} S_{\alpha \beta}(t) \right] 
\label{eq:Brouwer}
\end{equation}
where $S_{\alpha\beta}(t)$ is the scattering matrix at time $t$ from
channel $\alpha$ to channel $\beta$.  Here $S(t)$ is to be calculated
as if the parameters of the system are frozen at time $t$, and
$\alpha$ is summed only over channels at the source.  In the quantum
Hall regime, so long as there is no direct tunneling across the
quantum Hall bar (I.e, as long as the antidot is not simultaneously
connected to both edges), the structure of the scattering matrix is
trivial --- anything that comes into the left edge at the source
(bottom left of each frame of Fig.~\ref{fig:sequence}) must follow
that edge all the way to the drain (upper left).  If we have a quantum
Hall state with only a single edge channel ($\nu=1$, for example) the
scattering matrix has only two nonzero elements -- each with unit
magnitude (one element for the edge state leaving the source on the
lower left side and ending up at the upper left, and one leaving the
drain at the upper right and ending up at the source at the lower
right).  Only one of these two nonzero elements (the one representing
the state leaving the source) enters into Eq.  \ref{eq:Brouwer}.  We
write this relevant unit magnitude ($U(1)$ valued) element as $e^{i
  \phi(t)}$, such that we have the charge pumped per cycle as $Q = e
\int_0^\tau \frac{dt}{2 \pi} \,\,\, \frac{d \phi(t)}{dt} .  $ In the
integer quantum Hall regime the system must return to its original
state after a full cycle.  Thus, $\phi(t)$ must return to its original
value modulo $2 \pi$.  The pumped charge is then just the number of
times $\phi$ wraps by $2 \pi$ per cycle.  In this way we see that the
pumped charge is quantized as a result of being a topological
quantity!

This quantization argument can be generalized to the case of $m$
copropagating channels per edge.  In this case, the $m$ edge channels
can mix with each other as long as they all go directly along the edge
from the source to the drain and do not cross the Hall bar.  The
relevant nonzero terms of the scattering matrix then form a $U(m) =
U(1) \otimes SU(m)$ matrix.  It can be shown that the $U(1)$ part is
again the only important piece (representing the total charge) and the
pumped charge per cycle is again quantized as described above.

This scattering matrix formalism is easily extended to finite
temperature\cite{Brouwer} (at least for the integer case).  One needs
only to define scattering matrices $S(E,t)$ as a function of incoming
energy.  Eq.  \ref{eq:Brouwer} become $E$
dependent resulting in a charge transfer $Q(E)$ which is then smeared
by a Fermi function to give the charge transfer: $
  Q = \int \! dE \, Q(E) \,\, \frac{d n_F(E)}{dE} $
with $n_F$ the Fermi function.   In Fig.~\ref{fig:steps} this smearing 
by a Fermi function is shown as the dashed line (in the figure $T$ is
taken to be $10\%$ of the antidot single particle addition energy).

For the noninteracting electron (integer) case and for some simple
interacting cases, it is possible to solve for the scattering matrix
explicitly (given the energies of eigenstates on the antidot and the
tunneling matrix elements as a function of time).  Indeed it can be
established, as claimed above, that the charge pumped per cycle at
$T=0$ is quantized and is equal to the difference in the charge on the
antidot between steps (a) and (d).  

To generalize this scattering matrix approach to the fractional
quantum Hall regime, we imagine connecting a fractional Hall sample to
integer Hall leads in a smooth fashion\cite{Trans}, so that one can
still ask about the scattering matrix for electrons injected into the
system.  Here, due to the above mentioned ground state
degeneracy\cite{Stone}, the system need not return to its original
state after a single pumping cycle.  In the case of having $q$
degenerate ground states, the system can cycle through the ground
states returning to the original state only after $q$ full periods of
pumping.  Thus, the pumped charge $Q$ in Eq.  \ref{eq:Brouwer} need
only be quantized in units of the electron charge after $q$ cycles,
so the pumped current per cycle is quantized in units of $e/q$.

{\bf Experiments:} This experiment can thus be used as a measurement
of the charge of the fractional quantum Hall quasiparticle. Although,
a number of previous works have measured the fractional charge of
quantum Hall quasiparticles\cite{Goldman,Pepper2,Frac,endnote2}, it is
quite possible that the currently proposed pumping experiment will be
the theoretically clearest measurement yet.  

The main experimental problem in carrying out this experiment appears
to be that temperature must be sufficiently low that the current steps
(see Fig~\ref{fig:steps}) are not too smeared out.  As discussed
above, this temperature scale is mostly determined by the single
(quasi)particle addition energy for the antidot.  It is thus quite
useful to note that this energy scale has in fact been measured for
several similar experimental systems in both the integer and
fractional regimes\cite{Sac,Pepper1,Goldman,Pepper2}.  Although the
precise addition energy depends on the particular sample in question,
the authors of Refs.  \onlinecite{Sac,Pepper1,Goldman,Pepper2} were
able to achieve addition energies on the order of several hundred mK
for both $\nu=1$ and $\nu=1/3$.  For the case of $\nu=2/5$, however,
this energy seems to be somewhat lower\cite{Goldman}, but may still be
high enough to successfully perform the proposed pumping experiment.

Another experimental issue is how fast can one pump the system and
expect to have the pumped charge quantized.  This somewhat subtle
issue is a subject of current research.  However, as estimates, one
can expect that the tunnelling time from the antidot to the edge
should set one time scale, the single particle addition energy sets
another time scale, and the dissipation time yet another time
scale.   It is quite safe to say that pumping at a rate slower than
all of these time scales will remain quantized.  The effects of
pumping faster will be discussed in a forthcoming paper.

{\bf Acknowledgments:} I am indebted to B. Spivak for encouraging me
to think about pumping in quantum Hall systems, and to C. Marcus for
encouraging me to turn these ideas into a paper.  Helpful
conversations with N. Zhitenev, R. de Piccioto, L. Levitov, J. K.
Jain, A. Moustakas, and C. Chamon are also acknowledged.

\end{multicols}


\begin{references}
\bibitem{endnote1} Strictly speaking there is spectral flow of the
  ground state, and the state with an electron transferred from the
  inner edge to the outer edge is not the true ground state, but is
  returned to the true ground state by transferring an electron
  through an external wire (or a current meter) back from the outer to
  the inner edge.
\bibitem{endnote2} An important issue that arises is the question of
  whether the experiment measures the quasiparticle charge or just the
  filling fraction\cite{Frac}.  At $\nu=1/3$, for example, these two
  quantities are the same, so any measurement at this filling fraction
  is ambiguous in this sense.  At $\nu=2/5$, however, the charge of
  the quasiparticle is $1/5$.  The proposed pumping experiment at
  $\nu=2/5$ should properly pump charge quantized in units of $e/5$.
\end{references}
\end{document}